\documentclass[aps,pra,reprint,groupedaddress,longbibliography]{revtex4-1}

\usepackage{graphicx}
\usepackage{graphics}
\usepackage{amssymb}
\usepackage{natbib}

\begin{document}

\def\crta{\vrule height1.41ex depth-1.27ex width0.34em}
\def\dj{d\kern-0.36em\crta}
\def\Crta{\vrule height1ex depth-0.86ex width0.4em}
\def\Dj{D\kern-0.73em\Crta\kern0.33em}
\dimen0=\hsize \dimen1=\hsize \advance\dimen1 by 40pt

\newtheorem{definition}{Definition}[section]
\newtheorem{theorem}{Theorem}[section]
\newtheorem{corollary}{Corollary}[theorem]  
\newtheorem{lemma}{Lemma}[section]
\newtheorem{proposition}{Proposition}[section]

\title{Classical and Quantum Logics with Multiple and a Common
  Lattice Models}

\author{Mladen Pavi\v ci\'c}
\email{pavicic@irb.hr}
\homepage{http://www.irb.hr/users/mpavicic}
\affiliation{Department of Physics---Nanooptics,
Faculty of Mathematics and Natural Sciences,
Humboldt University of Berlin, Germany and\\
Center of Excellence for Advanced Materials 
and Sensing Devices (CEMS),\\ Photonics and Quantum Optics Unit, 
Ru{\dj}er Bo\v skovi\'c Institute, 
Zagreb, Croatia}


\begin{abstract}
We consider a proper propositional quantum logic and show that it 
has multiple disjoint lattice models, only one of which is an 
orthomodular lattice (algebra) underlying Hilbert (quantum) space. 
We give an equivalent proof for the classical logic which turns out
to have disjoint distributive and non-distributive ortholattices as
its models. In particular, we prove that quantum as well as classical
logics are complete and sound with respect to these lattices. We also
show that there is one common non-orthomodular lattice that is a model
of both quantum and classical logics. In technical terms, that enables
us to run the same classical logic on both a digital (standard, two
subset, 0-1 bit) computer and on a non-digital (say, a six subset)
computer (with appropriate chips and circuits). With quantum logic, 
the same six element common lattice can serve us as a benchmark 
for an efficient evaluation of equations of bigger lattice models or
theorems of the logic. 
\end{abstract}



\maketitle

\section{Introduction: Is Logic Empirical?}
\label{intro}

In his seminal paper {\em Is Logic Empirical?}~\cite{putnam-69},
Hilary Putnam argues that logic we make use of to handle the 
statements and propositions of the theories we employ to describe
the world around us is uniquely determined by it. 
``{\em Logic is empirical\/}. It makes \dots\ sense to speak of 
`physical logic'. We live in a world with a non-classical logic 
[of subspaces of the quantum Hilbert space $\mathcal{H}$ which form an 
orthomodular (non-distributive, non-Boolean) lattice]. Certain 
statements---just the ones we encounter in daily life---do obey 
classical logic, but this is so because the corresponding subspaces 
of $\mathcal{H}$ form a Boolean lattice.'' \cite[Ch.~V]{putnam-69} 

We see that Putnam, in effect, reduces the logics to lattices, 
while they should only be their models. ``[We] just read the 
logic off from the Hilbert space
$\mathcal{H}$.''~\cite[Ch.~III]{putnam-69} 
This technical approach has often been adopted in both classical and 
quantum logics. In classical logic, it has been known as two-valued 
interpretation for more than a century. In quantum logic, since 
G.~Birkhoff and J.~von Neumann introduced it in 1935
\cite{birk-v-neum} and it is still embraced by many authors
\cite{ql-stanford-enc-12}. Subsequently, varieties of relational logic
formulations, which closely follow lattice ordering relations, have
been developed, e.g., by Dishkant \cite{dishk-ol}, 
Goldblatt \cite{gold74}, Dalla Chiara \cite{dalla-c-h-b}, Nishimura 
\cite{nishimura80,nishimura-09}, Mittelstaedt 
\cite{mittelstaedt-book}, Stachow \cite{stachow-completness}, and 
Pt{\'a}k and Pulmannov{\'a} \cite{ptak-pulm}. 
More recently, Engesser and Gabbay \cite{engesser-gabay} made 
related usage of non-monotonic consequence relation, 
Rawling and Selesnick \cite{rawling} of binary sequent, 
Herbut \cite{herbut-12} of state-dependent implication of 
lattice of projectors in the Hilbert space, Tylec and Ku{\'s} 
\cite{tylec} of partially ordered set (poset) map, Bikchentaev, 
Navara, and Yakushev \cite{bi-navara-ya} of poset binary relation.

Another version of Birkhoff-von-Neumann style of viewing 
propositions as projections in Hilbert space rather than closed
subspaces and their lattices as in the original Birkhoff-von 
Neumann paper has been introduced by Engesser, Gabbay, and Lehmann
\cite{gabbay-quantum-book}. Recently, other versions of quantum 
logics have been developed, such as a dynamic quantum logic by 
Baltag and Smets \cite{baltag-smets-05,baltag-smets-11}, 
exogenous quantum propositional logic by Mateus and Sernadas 
\cite{mateus-sernadas-06}, a categorical quantum logic by 
Abramsky and Duncan \cite{abramsky-duncan-06,abramsky-coecke-04}, 
and a projection orthoalgebraic approach to quantum logic by
Harding \cite{harding-09}.

However, we are interested in non-relational logics which 
combine propositions according to a set of true formulas/axioms and 
rules imposed on them. The propositions correspond to statements from
a theory, say classical or quantum mechanics, and are not directly 
linked to particular measurement values. Such logics employ models 
which evaluate a particular combination of propositions and tell us 
whether it is true or not. Evaluation means mapping from a set of 
logic propositions to an algebra, e.g., a lattice, through which 
a correspondence with measurement values emerges, but indirectly. 
Therefore we shall consider a classical and a quantum logic defined 
as a set of axioms whose Lindenbaum-Tarski algebras of equivalence 
classes of expressions from appropriate lattices correspond to the 
models of the logics. Let us call such a logic an {\em axiomatic 
logic\/}.

We show that an axiomatic logic is wider than its relational logic
variety in the sense of having many possible models and not only
distributive ortholattice (Boolean algebra) for the classical logic
and not only orthomodular lattice for the quantum logic. We make use of 
Hilbert-Ackermann's presentation \cite{hilb-ack-book} of axiomatic
classical logic in the schemata form and of Kalmbach's axiomatic
quantum logic \cite{kalmb83,kalmb74} (in Megill-Pavi\v ci\'c
\cite{mpcommp99,pm-ql-l-hql1} presentation, i.e., without Kalmbach's
A1,A11 \&\ A15 axioms which we prove redundant in \cite{mphpa98}), as
typical examples of axiomatic logics. 

It is well-known that there are many interpretations of the 
classical logic, e.g., two-valued, general Boolean algebra 
(distributive ortholattice), set-valued ones, 
etc.~\cite[Ch.~8,9]{Schechter}. These different interpretations 
are tantamount to different models of the classical logic and in 
this paper and several previous papers of ours we show that they are 
enabled by different definitions of the relation of equivalence for 
its different Lindenbaum-Tarski algebras. One model of the classical 
logic is a distributive numerically valued, mostly two-valued, 
lattice, while the others are non-distributive non-orthomodular 
lattices, one of them being the so-called O6 lattice,  
which can also be given set-valuations \cite[Ch.~8,9]{Schechter}. 

As for quantum logic, one of its models is an orthomodular lattice, 
while others are non-orthomodular lattices, one of them being 
again O6---the common model of both logics. 

Within a logic we establish a unique deduction of all logic 
theorems from valid algebraic equations in a model and vice versa 
by proving the soundness and completeness of logic with respect 
to a chosen model. That means that we can infer the distributivity
or orthomodularity in one model and disprove them in another by 
means of the same set of logical axioms and theorems. We can also 
consider O6 in which both the distributivity and orthomodularity 
fail; however, particular non-distributive and non-orthomodular 
conditions pass O6 only to map into the distributivity and 
orthomodularity through classical and quantum logics in other 
models of these logics. 

We see that logic is at least not {\em uniquely\/} empirical 
since it can simultaneously describe distinct realities. 

The paper is organised as follows. In Sec.~\ref{sec:logic} we define 
classical and quantum logics. In Sec.~\ref{sec:latt} we introduce 
distributive (ortho)lattices and orthomodular lattices as well as 
two non-distributive (one is O6) and four non-orthomodular ones 
(one is again O6), all of which are our models for classical and 
quantum logic, respectively. In Subec.~\ref{subsec:sound}, we prove 
soundness and in Subec.~\ref{subsec:complete} completeness of 
classical and quantum logics with respect to the models introduced 
in Sec.\ref{sec:latt}. In Sec.~\ref{sec:conclusion}, we discuss the 
obtained results. 

\section{Logics}
\label{sec:logic}
We consider logic ($\mathcal{L}$) to be a language which is defined
by a set of conditions (axioms) and rules imposed on propositions.  
axioms and the rules of inference. We shall consider quantum 
as well as classical axiomatic logics.

The propositions in our axiomatic logic ($\mathcal{L}$) are well-formed 
formulae (wffs), which we define as follows.

Primitive (elementary) propositions are denoted as
$p_0,p_1,p_2,...$; primitive connectives are
$\neg$ (negation) and $\vee$ (disjunction).
$p_j$ is a wff for $j=0,1,2,...$; $\neg A$ is a wff if $A$ is a wff;
$A\vee B$ is a wff if $A$ and $B$ are wffs.

We define operations as follows.

\begin{definition}\label{D:conj}(Conjunction)
$$\qquad A\wedge B\ {\buildrel\rm def\over =}\ \neg (\neg A\vee\neg B).$$
\end{definition}

\begin{definition} \label{def:impl-0} (Classical implication)
$$\qquad A\to_{\rm c} B\ {\buildrel\rm def\over =}\ \neg A\vee B.$$
\end{definition}

\begin{definition} \label{def:impl-3} (Kalmbach's implication)
$$\qquad A\to_3 B\ \ {\buildrel\rm def\over =} \ \
(\neg A\wedge \neg B)\vee(\neg A\wedge B)\vee(A\wedge(\neg A\vee B)).\quad$$
\end{definition}

\begin{definition}\label{L:equiv} (Quantum equivalence)
$$\qquad A\equiv_{q} B\ \
{\buildrel\rm def\over =}\ \ (\neg A\wedge\neg B)\vee(A\wedge B).$$
\end{definition}

\begin{definition}\label{L:equiv-0} (Classical Boolean equivalence)
$$\qquad A\equiv_{\rm c} B\ \
{\buildrel\rm def\over =}\ \ (B\to_{\rm c} A)\wedge(A\to_{\rm c} B).$$
\end{definition}

Connectives bind in the following order
$\to$, $\equiv$, $\vee$, $\wedge$, $\neg$,
from weakest to strongest.

\smallskip
$\mathcal{F}^\circ$ is a set of all wffs. Algebra 
${\mathcal{F}}=\langle {\mathcal{F}}^\circ,\neg,\vee\rangle$
is built within $\mathcal{L}$ with wffs containing $\vee$ and $\neg$
by means of a set of axioms and rules of inference. In that way we get
other expressions called theorems (axioms are also theorems). We make
use of symbol $\vdash$ to denote the set of theorems; therefore
$A\in\ \vdash$ means that $A$ is a theorem; it can also be written as
$\vdash A$. We read it as: ``$A$ is provable,'' to mean: if $A$ is a
theorem, then there is a proof of it. We present our systems of axioms
in the schemata form, so that we do not have to make use of the rule
of substitution.

\begin{definition}\label{D:gamma-ql}
For\/ $\Gamma\subseteq{\mathcal{F}}^\circ$, $A$ is derivable from\/ 
$\Gamma$: $\Gamma\vdash_\mathcal{L} A$ (or simply\/ $\Gamma\vdash A$)
if there is a finite sequence of wffs, the last one of which is $A$,
and every one of them is either an axiom of $\mathcal{L}$ or a member
of\/ $\Gamma$ or obtained from its precursors by means of the rule of
inference of $\mathcal{L}$.
\end{definition}

\subsection{Classical Logic}
\label{subsec:cl-logic}

In the classical logic $\mathcal{CL}$, the sign 
$\vdash_{\mathcal{CL}}$ denotes provability from the axioms via the
rule of inference. We shall drop the subscript when it follows from
context, e.g., in the following axioms and the rule of inference that
define $\mathcal{CL}$.

{\bf Axioms}
\vskip-15pt\begin{eqnarray}
{\rm A1}\qquad &&\vdash A\vee A\to_c A\label{eq:cl-a1}\\
{\rm A2}\qquad &&\vdash A\to_c B\vee A\\
{\rm A3}\qquad &&\vdash B\vee A\to_c A\vee B\\
{\rm A4}\qquad &&\vdash (A\to_c B)\to_c(C\vee A\to_c C\vee B)\label{eq:cl-a4}
\end{eqnarray}
{\bf Rule of Inference} (traditionally called {\em Modus Ponens})
\vskip-15pt\begin{eqnarray}
{\rm R1}\qquad &&\vdash A \qquad \&\qquad
A\to_c B\qquad\Rightarrow\qquad\vdash B\label{eq:cl-r1}
\end{eqnarray}

No particular valuations of the primitive propositions wffs consist of
are assumed. We are only interested in whether wffs that are valid,
i.e., true under all possible valuations of the underlying models. We
show that the wffs that can be inferred from the axioms by means of
the rule inference are exactly those that are valid by proving the
soundness and completeness of the logic. 

\subsection{Quantum Logic}
\label{subsec:q-logic}

Quantum logic ($\mathcal{QL}$) 
is defined as a language consisting of propositions and 
connectives (operations) as introduced above, and the following 
axioms and a rule of inference. We will use $\vdash_{\mathcal{QL}}$ 
to denote provability from the axioms and the rule of 
$\mathcal{QL}$ and omit the subscript when it is obvious from 
the context, e.g., in the list of axioms and the rule of inference 
that follow.

\smallskip
\noindent{\bf Axioms}
\begin{eqnarray}
{\rm A2}\qquad &&\vdash A\equiv_{q} B\rightarrow_c
(B\equiv_{q} C\rightarrow_c
     A\equiv_{q} C)\label{eq:kalmb-A1}\\
{\rm A3}\qquad &&\vdash A\equiv_{q} B\rightarrow_c
\neg A\equiv_{q} \neg B\\
{\rm A4}\qquad &&\vdash A\equiv_{q} B\rightarrow_c
A\wedge C\equiv_{q} B\wedge C\\
{\rm A5}\qquad &&\vdash A\wedge B\equiv_{q} B\wedge A\\
{\rm A6}\qquad &&\vdash A\wedge (B\wedge C)\equiv_{q} 
(A\wedge B)\wedge C\\
{\rm A7}\qquad &&\vdash A\wedge (A\vee B)\equiv_{q} A\\
{\rm A8}\qquad &&\vdash\neg A\wedge A\equiv_{q}(\neg A\wedge 
A)\wedge B \\
{\rm A9}\qquad &&\vdash A\equiv_{q}\neg\neg A\\
{\rm A10}\qquad &&\vdash\neg(A\vee B)\equiv_{q}\neg A\wedge\neg B\\
{\rm A12}\qquad &&\vdash (A\equiv_{q} B)\equiv_{q}(B\equiv_{q} A)\\
{\rm A13}\qquad &&\vdash A\equiv_{q} B\rightarrow_c
(A\rightarrow_c B)\\
{\rm A14}\qquad &&\vdash (A\to_cB)\to_3
(A\to_3(A\to_3B))\label{eq:kalmb-A15}
\end{eqnarray}
{\bf Rule of Inference}
\begin{eqnarray}
\ {\rm R1}\qquad &&\vdash A \quad \& \quad \vdash A \rightarrow_3 B
\quad\Rightarrow \quad \vdash B\label{eq:kalmb-R1}
\end{eqnarray}

Soundness and completeness for quantum logic we prove below show that
the theorems which can be inferred from A1-14 via R1 are exactly those
that are valid.

\section{Lattices}
\label{sec:latt}

For the presentation of the main result it would be pointless and 
definitely unnecessary complicated to work with the full-fledged
models, i.e., Hilbert space, and the new non-Hilbert models that 
would be equally complex. It would be equally too complicated to 
present complete quantum or classical logic of the second order 
with all the quantifiers.  Instead, we shall deal with lattices and 
the propositional logics we introduced in Sec.~\ref{sec:logic}.
We start with a general lattice which contains all the other lattices 
we shall use later on. The lattice is called an {\em ortholattice} 
and we shall first briefly present how one arrives at it starting 
with Hilbert space. 

A Hilbert lattice is a kind of orthomodular lattice (see
Def.~\ref{mixed-id-q}).  In it the operation \it meet\/\rm, $a\cap b$,
corresponds to set intersection,
${\mathcal H}_a\bigcap{\mathcal H}_b$ of subspaces 
${\mathcal H}_a$ and ${\mathcal H}_b$  of the Hilbert space 
${\mathcal H}$; the ordering relation $a\le b$ corresponds to 
${\mathcal H}_a\subseteq{\mathcal H}_b$; the operation \it join\/\rm, 
$a\cup b$, corresponds to the smallest closed subspace of 
$\mathcal H$ containing ${\mathcal H}_a\bigcup{\mathcal H}_b$; 
and the \it orthocomplement\/\rm\ $a'$ corresponds
to ${\mathcal H}_a^\perp$, the set of vectors orthogonal to all vectors 
in ${\mathcal H}_a$. Within the Hilbert space there is the operation 
${\mathcal H}_a+{\mathcal H}_b$ (sum of two subspaces); it is defined as
as the set of sums of vectors from ${\mathcal H}_a$ and
${\mathcal H}_b$ but it has no a parallel in the Hilbert lattice.
The following ${\mathcal H}_a+{\mathcal H}_a^\perp={\mathcal H}$ holds.

One can define all the lattice operations on the Hilbert space itself
following the above definitions (${\mathcal H}_a\cap{\mathcal H}_b=
{\mathcal H}_a\bigcap{\mathcal H}_b$, etc.). Thus we have 
${\mathcal H}_a\cup{\mathcal H}_b=\overline{{\mathcal H}_a+
{\mathcal H}_b}=({\mathcal H}_a+{\mathcal H}_b)^{\perp\perp}=
({\mathcal H}_a^\perp\bigcap{\mathcal H}_b^\perp)^\perp$,
\cite[p.~175]{isham} where $\overline{{\mathcal H}_c}$ is the closure
of ${\mathcal H}_c$, and therefore ${\mathcal H}_a+
{\mathcal H}_b\subseteq{\mathcal H}_a\cup{\mathcal H}_b$.
For a finite dimensional ${\mathcal H}$ or for the orthogonal closed
subspaces ${\mathcal H}_a$ and  ${\mathcal H}_b$ we have
${\mathcal H}_a+{\mathcal H}_b={\mathcal H}_a\cup{\mathcal H}_b$.
\cite[pp.~21-29]{halmos}, \cite[pp.~66,67]{kalmb83},
\cite[pp.~8-16]{mittelstaedt-book}.

For vector $x\in{\mathcal H}$ that has a unique decomposition
$x=y+z$ for $y\in{\mathcal H}_a$ and $z\in{\mathcal H}_a^\perp$ there
is a projection $P_a(x)=y$ associated
with ${\mathcal H}_a$. The closed subspace which belong to $P$ is
${\mathcal H}_P=\{x\in {\mathcal H}|P(x)=x\}$. Let $P_a\cap P_b$ denote
a projection on ${\mathcal H}_a\cap{\mathcal H}_b$, $P_a\cup P_b\>$
a projection on ${\mathcal H}_a\cup{\mathcal H}_b$, $P_a+P_b\>$ a
projection on ${\mathcal H}_a+{\mathcal H}_b\>$ if ${\mathcal H}_a\perp
{\mathcal H}_b$, and let $P_a\le P_b$ means
${\mathcal H}_a\subseteq{\mathcal H}_b$. Then $a\cap b$ corresponds 
to $P_a\cap P_b=\lim_{n\to\infty}(P_aP_b)^n
$,\cite[p.~20]{mittelstaedt-book} $a'$ to $I-P_a$,
$a\cup b$ to $P_a\cup P_b=I-\lim_{n\to\infty}[(I-P_a)(I-P_b)]^n
$,\cite[p.~21]{mittelstaedt-book}  and $a\le b$ to  $P_a\le P_b$.
$a\le b$ also corresponds to either
$P_a=P_aP_b$ or to $P_a=P_bP_a$ or to $P_a-P_b=P_{a\cap b'}$.
Two projectors commute iff their associated closed subspaces commute.
This means that $a\cap(a'\cup b)\le b$
corresponds to $P_aP_b=P_bP_a$. In the latter case we have:
$P_a\cap P_b=P_aP_b$ and $P_a\cup P_b=P_a+P_b-P_aP_b$.
$a\perp b$, i.e., $P_a\perp P_b$ is characterised by $P_aP_b=0$.
\cite[pp.~173-176]{isham}, \cite[pp.~66,67]{kalmb83},
\cite[pp.~18-21]{mittelstaedt-book}, \cite[pp.~47-50]{holl70},

Closed subspaces ${\mathcal H}_a$,${\mathcal H}_b$,\dots\ as well 
as the corresponding projectors $P_a$,$P_b$,\dots\  form 
an algebra called the Hilbert lattice which is an 
ortholattice. The conditions of the following definition 
can be easily read off from the properties of the aforementioned
Hilbert subspaces or projectors. 

\begin{definition}\label{def:ourOL}
An {\em ortholattice} ({\rm OL\/}) is an algebra
$\langle{\mathcal{OL}}_0,',\cup,\cap\rangle$ in which
for any $a,b,c\in \,{\mathcal{OL}}_0$ {\rm \cite{mpqo02}}
the following conditions hold 
\begin{eqnarray}
&&a''\>=\>a\label{eq:notnot}\\
&&a\cup (a\cap b)\>=\>a\label{eq:ababa}\\
&&a\cup b\>=\>b\cup a\label{eq:aub}\\
&&a\cap b\>=\>(a'\cup b\,')'\label{eq:aAb}\\
&&a\cup (b\cup b\,')\>=\>b\cup b\,'\\
&&(a\cup b)\cup c\>=\>a\cup (b\cup c)\label{eq:aAbc}
\end{eqnarray}
Since $b\cup b'=a\cup a\,'$ for any $a,b\in
\,{\mathcal{OL}}_0$, we define the \/{\em least}\/ and the
\/{\em greatest}\/ elements of the lattice: 
\begin{eqnarray}
\rm\textstyle{0}\,{\buildrel\rm def\over =}a\cap a',\qquad\qquad
\qquad\textstyle{1}\,{\buildrel\rm def\over=}a\cup a',
\label{eq:onezero}
\end{eqnarray}
\parindent=0pt
and the {\rm ordering relation ($\le$) on the lattice}:
\begin{eqnarray}
  \qquad a\le b\ \quad{\buildrel\rm def\over\Longleftrightarrow}\quad\
  a\cap b=a \quad\Longleftrightarrow\quad a\cup b=b,\qquad
\end{eqnarray}
\end{definition}

\parindent=20pt

\begin{definition}\label{def:impl-L} (Sasaki hook)
\begin{eqnarray}
a\to_1b\ \ {\buildrel\rm def\over =}\ \
a'\cup(a\cap b)
\end{eqnarray}
\end{definition}

\begin{definition}\label{L:id-bi-L-q} (Quantum equivalence) 
\begin{eqnarray}
a\equiv_q b\ \
{\buildrel\rm def\over =}\ \ (a\cap b)\cup(a'\cap b\,').
\end{eqnarray}
\end{definition}

\begin{definition}\label{L:id-bi-L-c} (Classical equivalence) 
\begin{eqnarray}
a\equiv_c b\ \
{\buildrel\rm def\over =}\ \ (a'\cup b)\cap(b\,'\cup a).
\end{eqnarray}
\end{definition}

Connectives bind in the following order
$\to$, $\equiv $, $\cup$, $\cap$, and $'$,
from weakest to strongest.

\begin{definition}\label{mixed-id-q}{\em (Pavi\v ci\'c, \cite{pav93}.)}
An orthomodular lattice {\em (OML)\/} is an {\em OL\/} in which 
condition (\ref{eq:qm-as-id-q}) (called {\em orthomodularity\/}) holds
\begin{eqnarray}
a\equiv_q b=1\qquad \Rightarrow\qquad a=b.\label{eq:qm-as-id-q}
\end{eqnarray}
\end{definition}

Every Hilbert space (finite and infinite) and every phase space 
is orthomodular. 

\begin{definition}\label{mixed-id-c} {\em (Pavi\v ci\'c, 
\cite{p98}.)\footnote{The proof of the opposite claim in 
\cite[Theorem 3.2]{pav93} is wrong.}}
A distributive ortholattice {\em (DL)\/} (also called a 
Boolean algebra) is an \/{\em OL\/} in which condition
(\ref{eq:qm-as-id-c}) (called {\em distributivity}) holds
\begin{eqnarray}
a\equiv_c b=1\qquad \Rightarrow\qquad a=b.\label{eq:qm-as-id-c}
\end{eqnarray}
\end{definition}

Every phase space is distributive and, of course, orthomodular 
since every distributive ortholattice is orthomodular.

The opposite directions of meta-implications in 
Eqs.~(\ref{eq:qm-as-id-q}) and (\ref{eq:qm-as-id-c}) hold in any OL.

\begin{definition}\label{def:woml} {\em (Pavi\v ci\'c and Megill,  
\cite{mpcommp99}.)} A {\em weakly orthomodular ortholattice}
{\rm (WOML)} is an {\em OL} in which either of conditions
(\ref{eq:woml},\ref{eq:woml-a}) (called {\em weak orthomodularity\/})
hold
\begin{eqnarray}
a\to_1b=\textstyle{1}\qquad&\Rightarrow&\qquad 
b'\to_1a'=\textstyle{1},\label{eq:woml}\\
a\equiv  b=1\qquad&\Rightarrow&\qquad
(a\cup c)\equiv(b\cup c)=1.\qquad
\label{eq:woml-a}
\end{eqnarray}
\end{definition}

\begin{definition}\label{def:woml1} {\em (Pavi\v ci\'c, this paper.)\/} 
A \/{\rm WOML1}\/ is a \/{\rm WOML}\/ in which 
\begin{eqnarray}
[(a\to_1b)\equiv(b\to_1a)]\ =\ (a\equiv b)
\label{eq:woml1}
\end{eqnarray}
holds.
\end{definition}

\begin{definition}\label{def:woml2} {\em (Pavi\v ci\'c, this paper.)\/} 
A \/{\rm WOML2}\/ is a \/{\rm WOML1}\/ in which
\begin{eqnarray}
[(a\equiv b)'\to_1a']\ =\ (a\to_1b)
\label{eq:woml2}
\end{eqnarray}
holds.
\end{definition}

\begin{definition}\label{def:woml*} {\em (Pavi\v ci\'c, this paper.)\/} 
A \/{\rm WOML*}\/ is a \/{\rm WOML}\/ in which neither
Eq.~(\ref{eq:qm-as-id-q}), nor (\ref{eq:woml2}), nor
(\ref{eq:woml1}) hold. 
\end{definition}

\begin{definition}\label{def:woml3} {\em (Pavi\v ci\'c and Megill,  
\cite{mpcommp99,pmjlc08}.)} A {\em weakly distributive ortholattice}
{\rm (WDL)} is an {\em OL} in which condition (\ref{eq:wdl-a3})
(called {\em commensurability\/}) holds
\begin{eqnarray}
(a\cap b)\cup(a\cap b')\cup(a'\cap b)\cup(a'\cap b')=1.
\label{eq:wdl-a3}
\end{eqnarray}
\end{definition}

\begin{definition}\label{def:woml4} {\em (Pavi\v ci\'c and Megill,  
\cite{mpcommp99}.)} A {\em weakly distributive ortholattice}
{\rm (WDL)} is a {\em WOML} in which condition (\ref{eq:wdl-a4}) 
(called {\em weak distributivity}) holds
\begin{eqnarray}
a\cup(b\cap c)\equiv_c(a\cup b) \cap (a\cup c)=1.
\label{eq:wdl-a4}
\end{eqnarray}
\end{definition}

Definitions \ref{def:woml3} and \ref{def:woml4} are equivalent. 
We give both definitions here in order to, on the one hand, stress 
that a WDL is a lattice in which all variables are commensurable 
and, on the other, to show that in WDL the distributivity holds 
only in its weak form given by Eq.~(\ref{eq:wdl-a4}) which we will 
use later on. 

\begin{definition}\label{def:wdl*} {\em (Pavi\v ci\'c and Megill,  
\cite{pmjlc08})\/} A \/{\rm WDL*}\/ is a \/{\rm WDL}\/ in which
Eq.~(\ref{eq:qm-as-id-c}) does not hold. 
\end{definition}

\begin{figure}[htbp]\centering
\begin{picture}(320,200)(-10,-10)
 \put(-5,100){
      \begin{picture}(40,80)(0,0)
    \put(20,0){\line(-1,1){20}}
    \put(20,0){\line(1,1){20}}
    \put(0,20){\line(0,1){20}}
    \put(40,20){\line(0,1){20}}
    \put(0,40){\line(1,1){20}}
    \put(40,40){\line(-1,1){20}}

    \put(20,-5){\makebox(0,0)[t]{$\textstyle{0}$}}
    \put(-5,20){\makebox(0,0)[r]{$x$}}
    \put(45,20){\makebox(0,0)[l]{$y'$}}
    \put(-5,40){\makebox(0,0)[r]{$y$}}
    \put(45,40){\makebox(0,0)[l]{$x'$}}
    \put(20,65){\makebox(0,0)[b]{$\textstyle{1}$}}
    \put(15,75){\makebox(0,0)[b]{$\textstyle{(a)}$}}

    \put(20,0){\circle*{3}}
    \put(0,20){\circle*{3}}
    \put(40,20){\circle*{3}}
    \put(0,40){\circle*{3}}
    \put(40,40){\circle*{3}}
    \put(20,60){\circle*{3}}
  \end{picture} 
} 

    \put(8,0){
       \begin{picture}(100,100)(0,0)

    \put(50,0){\line(-1,1){40}}
    \put(50,0){\line(0,1){40}}
    \put(50,0){\line(1,1){40}}
    \put(10,40){\line(0,1){20}}
    \put(10,60){\line(2,1){40}}
    \put(50,40){\line(-1,1){40}}

    \put(50,40){\line(2,1){40}}
    \put(90,40){\line(-1,1){40}}
    \put(90,40){\line(0,1){20}}
    \put(10,60){\line(0,1){20}}
    \put(90,60){\line(0,1){20}}
    \put(10,80){\line(1,1){40}}
    \put(50,80){\line(0,1){40}}
    \put(90,80){\line(-1,1){40}}

    \put(50,-5){\makebox(0,0)[t]{$\textstyle{0}$}}
    \put(5,40){\makebox(0,0)[r]{$x$}}
    \put(50,45){\makebox(0,0)[b]{$w$}}
    \put(95,40){\makebox(0,0)[l]{$z'$}}
    \put(5,60){\makebox(0,0)[r]{$y$}}
    \put(95,60){\makebox(0,0)[l]{$y'$}}
    \put(5,80){\makebox(0,0)[r]{$z$}}
    \put(50,73){\makebox(0,0)[t]{$w'$}}
    \put(95,80){\makebox(0,0)[l]{$x'$}}
    \put(50,125){\makebox(0,0)[b]{$\textstyle{1}$}}
    \put(60,135){\makebox(0,0)[b]{$\textstyle{(b)}$}}
    \put(50,0){\circle*{3}}
    \put(10,40){\circle*{3}}
    \put(50,40){\circle*{3}}
    \put(90,40){\circle*{3}}
    \put(10,60){\circle*{3}}
    \put(90,60){\circle*{3}}
    \put(10,80){\circle*{3}}
    \put(50,80){\circle*{3}}
    \put(90,80){\circle*{3}}
    \put(50,120){\circle*{3}}

      \end{picture}
    } 

      \put(120,35) { 
        \begin{picture}(100,100)(0,0) 
          \put(0,25){\line(0,1){75}}
          \put(0,25){\line(2,-1){50}}
          \put(0,25){\line(1,1){50}}
          \put(0,50){\line(2,-1){50}}
          \put(0,50){\line(1,1){25}}
          \put(0,100){\line(1,-1){37.5}}
          \put(0,100){\line(2,1){50}}
          \put(50,0){\line(0,1){50}}
          \put(50,125){\line(0,-1){50}}
          \put(50,50){\line(-3,1){37.5}}
          \put(100,100){\line(0,-1){75}}
          \put(100,100){\line(-2,1){50}}
          \put(100,100){\line(-1,-1){50}}
          \put(100,75){\line(-2,1){50}}
          \put(100,75){\line(-1,-1){25}}
          \put(100,25){\line(-1,1){37.5}}
          \put(100,25){\line(-2,-1){50}}
          \put(50,75){\line(3,-1){37.5}}

          \put(0,25){\circle*{3}}
          \put(0,50){\circle*{3}}
          \put(0,75){\circle*{3}}
          \put(0,100){\circle*{3}}
          \put(12.5,62.5){\circle*{3}}
          \put(25,75){\circle*{3}}
          \put(37.5,62.5){\circle*{3}}
          \put(50,0){\circle*{3}}
          \put(50,25){\circle*{3}}
          \put(50,50){\circle*{3}}
          \put(50,75){\circle*{3}}
          \put(50,100){\circle*{3}}
          \put(50,125){\circle*{3}}
          \put(62.5,62.5){\circle*{3}}
          \put(75,50){\circle*{3}}
          \put(87.5,62.5){\circle*{3}}
          \put(100,25){\circle*{3}}
          \put(100,50){\circle*{3}}
          \put(100,75){\circle*{3}}
          \put(100,100){\circle*{3}}

          \put(0,25){\makebox(0,0)[r]{$w\ $}}
          \put(0,50){\makebox(0,0)[r]{$z\ $}}
          \put(0,75){\makebox(0,0)[r]{$y\ $}}
          \put(0,100){\makebox(0,0)[r]{$x\ $}}
          \put(12.5,57.5){\makebox(0,0)[t]{$\ t$}}
          \put(28,77){\makebox(0,0)[b]{$\ s$}}
          \put(37.5,67.5){\makebox(0,0)[b]{$r$}}
          \put(50,-5){\makebox(0,0)[t]{$\textstyle{0}$}}
          \put(50,25){\makebox(0,0)[l]{$\ v'$}}
          \put(50,47){\makebox(0,0)[t]{$u'\ \ \ $}}
          \put(50,78){\makebox(0,0)[b]{$\ \ \ u$}}
          \put(50,100){\makebox(0,0)[r]{$v\ $}}
          \put(50,130){\makebox(0,0)[b]{$\textstyle{1}$}}
         \put(45,140){\makebox(0,0)[b]{$\textstyle{(c)}$}}
          \put(62.5,57.5){\makebox(0,0)[t]{$r'$}}
          \put(70,50){\makebox(0,0)[t]{$s'\ $}}
          \put(87.5,67.5){\makebox(0,0)[b]{$t'\ $}}
          \put(100,25){\makebox(0,0)[l]{$\ x'$}}
          \put(100,50){\makebox(0,0)[l]{$\ y'$}}
          \put(100,75){\makebox(0,0)[l]{$\ z'$}}
          \put(100,100){\makebox(0,0)[l]{$\ w'$}}


        \end{picture}
      } 

  \end{picture}
\caption{(a) O6; (b)
O7 (Beran, Fig.~7b \cite{beran}); (c) O8 (Rose-\-Wilkinson-1 
\cite{mpijtp03b})\label{fig:womls}}
\end{figure}
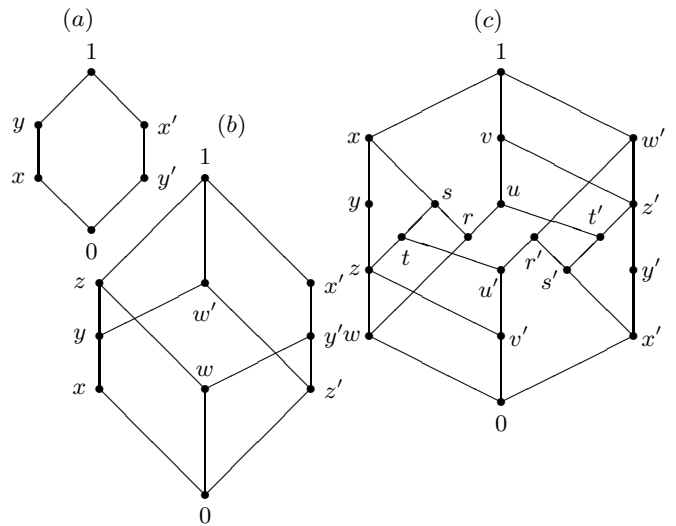

We represent finite lattices by a {\rm Hasse diagrams} which consist
of {\em vertices} (dots) and {\em edges} (lines that connect dots).
Each dot represents an element in a lattice, and positioning
an element $a$ above another element $b$ and connecting them by a line
means $a\le b$. E.g., in Fig.~\ref{fig:womls}~(a) we have
$0\le x\le y\le 1$. There, for instance, $x$ is not in a relation with
either $x'$ or $y'$.

The statement ``orthomodularity (\ref{eq:qm-as-id-q}) does not 
hold in WOML*''  reads $\sim\!\!\![(\forall a,b\in {\rm WOML}*)
((a\!\equiv\! b=1)\ \Rightarrow\ (a=b))]$ what we can write as 
$(\exists a,b\!\!\in\!\! {\rm WOML}*)(a\!\!\equiv\!\! b\!=\!1\ \&\ 
a\ne b)$, where ``$\sim$'' is a meta-negation and ``$\&$'' a
meta-conjuc\-tion. An example of a WOML* is O6 from 
Fig.~\ref{fig:womls}~(a) and we can easily check the statement on it. 
O6 is also an example of a WDL* and we can verify the statement 
``distributivity (\ref{eq:qm-as-id-c}) does not hold in WDL*'' on it, 
as well. Similarly, ``condition (\ref{eq:woml2}) does not
hold in WOML*'' we can write as $(\exists a,b\!\in\! {\rm WOML}*)(
((a\!\equiv\! b)'\to_1a')\ \ne\ (a\to_1b))$.

\begin{definition}\label{def:woml1*} {\em (Pavi\v ci\'c,
this paper.)\/} A \/{\rm WOML1*}\/ is a \/{\rm WOML1}\/ in which
neither Eq.~(\ref{eq:qm-as-id-q}) nor (\ref{eq:woml2}) hold. 
\end{definition}

An example of a WOML1* is O7 from Fig.~\ref{fig:womls}~(b).

\begin{definition}\label{def:woml2*} {\em (Pavi\v ci\'c,
this paper.)\/} A \/{\rm WOML2*}\/ is a \/{\rm WOML2}\/ in which
Eq.~(\ref{eq:qm-as-id-q}) does not hold.  
\end{definition}

An example of a WOML2* is O8 from Fig.~\ref{fig:womls}~(c).

\begin{lemma}\label{lemma:woml0-2} 
{\rm OML} is properly included in (i.e., it is stronger than) 
{\rm WOML2}, {\rm WOML2} is properly included in {\rm WOML1}, 
and {\rm WOML1} is properly included in {\rm WOML}, 

\end{lemma}
{\parindent=0pt{\em Proof.} Eq.~(\ref{eq:woml}) passes O6, O7, and O8 
from Fig.~\ref{fig:womls}. Eq.~(\ref{eq:woml1}) passes O7 and O8, but 
fails in O6. Eq.~(\ref{eq:woml2}) passes O8 but fails in 
both O6 and O7.  Eq.~(\ref{eq:qm-as-id-q}) fails in 
O6, O7, and O8. To find the failures and passes we used our 
program {\tt lattice} \cite{bdm-ndm-mp-1}.  
$\phantom .$\hfill$\blacksquare$}

\begin{lemma}\label{lemma:woml0-2*}
{\rm OML}\/ is included in neither {\rm WOML2*}\/, nor 
\/{\rm WOML1*}\/,  nor \/{\rm WOML*}\/.
{\rm WOML2*}\/ is included in neither {\rm WOML1*}\/,
nor \/{\rm WOML*}\/. {\rm WOML1*}\/ is not included in 
\/{\rm WOML*}\/.

\end{lemma}
{\parindent=0pt{\em Proof. } The proof straightforwardly follows from
the one of Lemma \ref{lemma:woml0-2} and the definitions of
{\rm WOML*}, {\rm WOML1*}, {\rm WOML2*}, and  {\rm OML}.
$\phantom .$\hfill$\blacksquare$}

According to Definitions \ref{def:woml*}, \ref{def:woml1*}, 
\ref{def:woml2*}, and \ref{def:wdl*},  of WOML*, WOML1*, WOML2*, 
and WDL*, respectively, these lattices denote set-theoretical 
differences and that is going to play a crucial role in our 
proofs of completeness in Subsection \ref{subsec:complete} 
in contrast to \cite{mpcommp99} where we considered only WOML
without excluding the orthomodular equation. In Subsection 
\ref{subsec:complete} we shall come back to this decisive 
difference between the two approaches. Note that the 
set-differences are not equational varieties. For instance, 
WOML2* is a WOML2 in which the orthomodularity condition does 
not hold, but we cannot obtain WOML2* from WOML2 by {\em adding\/} 
new equational conditions to those defining WOML2. Instead, 
WOML2* can be viewed as a set of lattices in all of which the 
orthomodularity condition is violated.   

\medskip
\parindent=0pt
{\em Remarks on implications\/}. As we could see above, the 
implications do not play any decisive role in the definition of 
lattices, especially not in the definitions of OML and DL where 
they do not appear at all, and they also do not play a decisive 
role in the definition of logics. A few decades ago that was 
a major issue, though: ``A `logic' without an implication
... is radically incomplete, and hardly qualifies
as a theory of deduction'' ~\cite{zeman78} and a hunt to find a
``proper implication'' among the five possible ones was pursued  
in 1970ies and 1980ies \cite{harde79,pav-bibl,mpijtp98}.  
Apart from $\to_1$ and $\to_3$ it turns out \cite{kalmb83} that one 
can also define $a\to_0 b\ {\buildrel\rm def\over =}\ a'\cup b$
(classical), $a\rightarrow_2b\ {\buildrel\rm def\over =}\ 
b'\rightarrow_1a'$ (Dishkant), $a\rightarrow_4\nobreak b\ 
{\buildrel\rm def\over =}\ b'\rightarrow_3a'$ (non-tollens), and  
$a\rightarrow_5b\ {\buildrel\rm def\over =}\ (a\cap b)\cup(a'\cap b)
\cup(a'\cap b')$ (relevance).
In 1987 Pavi\v ci\'c \cite{pav87} proved that an OL in which
we have $a\rightarrow_i b=1\ \Rightarrow\ a\le b,\ i=1,\dots,5$ is an
OML. In 1987 Pavi\v ci\'c \cite{pav87} also proved that an OL in which
we have $a\rightarrow_0b=1 \Rightarrow a\le\nobreak b$ is a DL.
Therefore 5 different but nevertheless equivalent relational
logics could be obtained by linking lattice inequality to 5 
implications. With our linking of a single equivalence to lattice 
equality this ambiguity is avoided and we obtain a uniquely defined 
axiomatic quantum logic. Note that we have 
$a\equiv_qb=(a\to_ib)\cap(b\to_ia),\ i=1,\dots,5$ 
in every OML but not in every OL. {\em End of remarks.}
\parindent=20pt

\section{Soundness and Completeness}
\label{sec:models}

We shall connect our logics with our lattices so as to show 
that the latter are the models of the former.  

\begin{definition}\label{exists-c}We call ${\mathcal{M}}=\langle
L,h\rangle$ a model if $L$ is an algebra and
$h:{\mathcal{F}}^\circ\longrightarrow L$, called a valuation,
is a morphism of formulae ${\mathcal{F}}^\circ$
into $L$, preserving the operations $\neg,\vee$
while turning them into $',\cup$.
\end{definition}

When $L$ belongs to O6, WOML*, WOML1*, WOML2*, OML, WDL*, or DL
we can informally say that the model belongs to O6, WOML*, \dots, DL.
So, when we say ``for all models in O6, WOML*, \dots, DL,'' that
means ``for all base sets in O6, WOML*, \dots, DL and for all
valuations on each base set.'' ``Model'' might refer to a particular
pair $\langle L,h\rangle$ or to all such pairs with the base set $L$,
as would follow from the context.

\begin{definition}\label{one}We call a formula $A\in{\mathcal{F}}^\circ$
valid in the model $\mathcal{M}$, and write $\vDash_{\mathcal{M}} A$, if
$h(A)=\textstyle{1}$ for all valuations $h$ on the model, i.e.,
for all $h$ associated with the base set $L$ of the model. 
We call a formula $A\in{\mathcal{F}}^\circ$ a consequence of\/ 
$\Gamma\subseteq{\mathcal{F}}^\circ$ in the 
model $\mathcal{M}$ and write $\Gamma\vDash_{\mathcal{M}} A$ 
if $h(X)=\textstyle{1}$ for all $X$ in\/ $\Gamma$ implies 
$h(A)=\textstyle{1}$, for all valuations $h$.
\end{definition}

\subsection{Soundness}
\label{subsec:sound}

Proving soundness means proving that the axioms and rules of inference
and consequently all theorems of $\mathcal{QL}$ hold in the models of
$\mathcal{QL}$. The models of ${\mathcal{QL}}$ are O6, WOML*, WOML1*,
WOML2*, and OML, and of ${\mathcal{CL}}$ are O6, WDL* and DL. With the 
exception of O6 which is a special case of both WOML* and WDL*, they do
not properly include each other. 

$\vDash_{\mathcal{M}} A$ and $\Gamma\vDash_{\mathcal{M}} A$ are
implicitly quantified over all appropriate lattice models
$\mathcal{M}$. Statement ``valid'' without qualification will mean
valid in all appropriate models.

The theorems \ref{th:soundness-c} and \ref{th:soundness-q} below show
that if $A$ is a theorem of ${\mathcal{QL}}$, then $A$ will be valid
in O6, and any WOML*, WOML1*, WOML2*, or OML model, and if $A$ is a
theorem of ${\mathcal{CL}}$, then $A$ will be valid in O6, and any WDL* 
or DL model. In \cite{mpcommp99,pm-ql-l-hql1} we proved the 
soundness for WOML. Since that proof uses no additional conditions 
that hold in O6, WOML*, \dots, OML the proof given there for WOML 
is a proof of soundness for O6, WOML*, WOML1*, WOML2*, and
OML, as well. Also, in \cite{mpcommp99,pm-ql-l-hql1} we proved 
the soundness for WDL. Since that proof uses no additional 
conditions that hold in O6, WDL* and DL, the proof given there 
for WDL is a proof of soundness for O6, WDL*, and DL, as well.
Hence, we can prove the soundness of quantum and classical logic 
with the help of WOML and WDL conditions without referring to 
conditions (\ref{eq:qm-as-id-c}),  (\ref{eq:qm-as-id-q}), 
(\ref{eq:woml2}), or (\ref{eq:woml1}), i.e., to any condition in 
addition to those that hold in the WOML and WDL themselves. 

\begin{theorem}\label{th:soundness-c}{\rm [Soundness of 
${\mathcal{CL}}$.]}
$$\quad\Gamma\vdash_{\mathcal{CL}} A\quad\Rightarrow\quad
\Gamma\vDash_{\rm WDL} A$$
\end{theorem}

{\parindent=0pt{\em Proof.}
By Theorem 4.3 of \cite{mpcommp99} any WDL 
(in particular, O6, WDL* or DL) is a model for ${\mathcal{CL}}$.  
$\phantom .$\hfill$\blacksquare$}

\begin{theorem}\label{th:soundness-q}{\rm [Soundness of
${\mathcal{QL}}$.]}
$$\quad\Gamma\vdash_{\mathcal{QL}} A\quad\Rightarrow\quad
\Gamma\vDash_{\rm WOML} A$$
\end{theorem}

{\parindent=0pt{\em Proof.}
By Theorem 3.10 of \cite{mpcommp99} any WOML 
(in particular, O6, WOML*, WOML1*, WOML2*, or OML)
is a model for ${\mathcal{QL}}$.  
$\phantom .$\hfill$\blacksquare$}

Theorems \ref{th:soundness-c} and \ref{th:soundness-q}
express the fact that $\quad\Gamma\vdash_{\mathcal{CL}} A$
and $\Gamma\vdash_{\mathcal{QL}} A$ in axiomatic logics
${\mathcal{CL}}$ and ${\mathcal{QL}}$ correspond to 
$a=h(A)=1$ in their lattice models, from O6 and WOML till 
WDL. That means that we do not arrive at equations of the 
form $a=b$ and that starting from $\quad\Gamma\vdash A\equiv_q B$
we cannot arrive at $a=h(A)=b=h(B)$ but only at $a\equiv_q b=1$.
We can obtain a better understanding of this through the following
properties of OML and DL. The equational theory of OML consists of
equality conditions Eqs.~(\ref{eq:notnot})--(\ref{eq:aAbc}) together
with the orthomodularity equality condition \cite{pm-ql-l-hql1}
\begin{eqnarray}
a\cup(a'\cap(a\cup b))=a\cup b\label{eq:oml2o1}
\end{eqnarray} 
which is equivalent to the condition given by 
Eq.~(\ref{eq:qm-as-id-q}).
We now map each of these OML equations, which are of the form 
$t=s$, to the form $t\equiv_q s=1$. This is possible in any WOML 
since 
\begin{eqnarray}
a\cup(a'\cap(a\cup b))\equiv_q a\cup b=1\label{eq:oml2o11}
\end{eqnarray} 
holds in every OL \cite{pm-ql-l-hql1} and 
Eqs.~(\ref{eq:notnot})--(\ref{eq:aAbc}) mapped to the form 
$t\equiv_q s=1$ also hold in any OL. Any equational proof in 
OML can then be simulated in WOML by replacing each axiom 
reference in the OML proof with its corresponding WOML 
mapping. Such mapped proofs will make use of just a proper 
subset of the equations that hold in WOML.

It follows that equations of the form $t\equiv_q s=1$, where 
$t$ and $s$ are such that $t=s$ holds in OML, cannot determine
OML when added to an OL since all such forms pass O6 and an 
OL is an OML if and only if it does not include a subalgebra 
isomorphic to O6 \cite{holl70}.  

As for ${\mathcal{CL}}$, the equational theory of distributive
ortholattices can be simulated by a proper subset of the 
equational theory of WDLs since it consists of equality conditions 
Eqs.~(\ref{eq:notnot})--(\ref{eq:aAbc}) together with the 
distributivity equation 
\begin{eqnarray}
a\cap(b\cup c)=(a\cap b)\cup(a\cap c)\label{eq:ba2ao}
\end{eqnarray} 
which is equivalent to the condition (\ref{eq:qm-as-id-c}).
As with WOML above, we map these algebra conditions of the form 
$t=s$  to the conditions of the form $t\equiv_c s=1$, which hold
in any WDL since the weak distributivity condition given by
Eq.~(\ref{eq:wdl-a4}) holds in any WDL. Any equational proof in a
DL can then be simulated in WDL by replacing each condition in a
DL proof with its corresponding WDL mapping. Such a mapped proof
will use only a proper subset of the equations that hold in WDL.

Therefore, no set of equations of the form $t\equiv_cs=1$, 
where $t=s$ holds in DL, can determine a DL when added to an OL.
Such equations hold in WDL and none of the WDL equations 
(\ref{eq:notnot})--(\ref{eq:aAbc},\ref{eq:ba2ao}) is violated by
O6 which itself violates the distributivity condition 
\cite{pm-ql-l-hql1}. 

Similar reasoning applies to O6, WOML*, WOML1, WOML1*, WOML2, 
and WOML2* which are all WOMLs and to O6 and WDL* which are WDLs. 
Soundness applies to them all through WOML and WDL and which 
particular model we shall use for $\mathcal{QL}$ and $\mathcal{CL}$ is 
determined by a particular Lindenbaum-Tarski algebra which 
we use for the completeness proofs in the next subsection.

\subsection{Completeness}
\label{subsec:complete}

The soundness of $\mathcal{CL}$ and $\mathcal{QL}$ in
Subsec.~\ref{subsec:sound} shows that axioms and rules of inference
and all theorems from $\mathcal{CL}$ and $\mathcal{QL}$ hold in any
WOML. The completeness of $\mathcal{CL}$ and $\mathcal{QL}$ shows the 
opposite, i.e., that we can impose the structures of O6, WDL* and DL, 
and O6, WOML*, WOML1*, WOML2*, and OML on the sets 
${\mathcal{F}}^\circ$ of formulae of $\mathcal{CL}$ and $\mathcal{QL}$, 
respectively. But here, as opposed to the soundness proof, we shall 
have as many completeness proofs as there are models. The 
completeness proofs for O6, WOML*,  WOML1*, and WOML2* can be inferred 
neither from the proof for OML nor from the proofs for the other three
models. The same holds for O6, WDL* and DL.  

To establish a correspondence between formulae of $\mathcal{QL}$ and 
$\mathcal{CL}$ and conditions of O6, WOML*, WOML1*, WOML2*, and OML, 
and O6, WDL*, and DL, respectively, we make use of an equivalence
relation compatible with the operations in $\mathcal{QL}$ and
$\mathcal{CL}$, i.e., a relation of congruence. The resulting
equivalence classes stand for elements of these lattices and enable
the completeness proof of $\mathcal{QL}$ and $\mathcal{CL}$ for them.

The definition of the congruence relation involves a special set of
valuations on lattices O6, O7, and O8 (see Fig.~\ref{fig:womls})
called ${\mathcal{O}}${\rm 6}, ${\mathcal{O}}${\rm 7}, and
${\mathcal{O}}${\rm 8}.

\begin{definition}\label{D:hexagon}Letting ${\rm Oi}$, {\rm i=6,7,8},
represent the lattices from Figure \ref{fig:womls}, we define 
${\mathcal{O}}${\rm i} as the set of all mappings 
$o_{\rm i}:{\mathcal{F}}^\circ\longrightarrow {\rm Oi}$ 
such that for $A,B\in{\mathcal{F}}^\circ$,
$o_{\rm i}(\neg A)=o_{\rm i}(A)'$, and 
$o_{\rm i}(A\vee B)=o_{\rm i}(A)\cup o_{\rm i}(B)$.
\end{definition}

${\mathcal{O}}${\rm i}, ${\rm i}=6,7,8$ enable us to distinguish the
equivalence classes used for the completeness proof, so that the
Lindenbaum-Tarski algebras be O6, WOML*, WOML1*, and WOML2*.

We achieve that by conjoining the term $(\forall o_{\rm i}\in
{\mathcal{O}}{\rm i})\{[(\forall X\in\Gamma)(o_{\rm i}(X)=
\textstyle{1})]\Rightarrow [o_{\rm i}(A)=o_{\rm i}(B)]\}$, i=6,7,8,
to the definition of the equivalence relation so that the valuations of
wffs $A$ and $B$ map to the same point in the lattice Oi whenever the
valuations $o_{\rm i}$ of the wffs in $\Gamma$ are all $\textstyle{1}$.
Thus, e.g., in O6 wffs $A\vee B$ and $A\vee (\neg A\wedge (A\vee B))$
become members of two separate equivalence classes, what by Theorem
\ref{th:non-distr} below, amounts to non-orthomodularity of WOML. If
it were not for the conjoined term, the two wffs would belong to the
same equivalence class. Conjoined terms provide a completeness proof
that is not in any way dependent on the orthomodular law. Therefore to
prove the completeness the underlying models need not be orthomodular.
The equivalence classes so defined work for WOML1*, and WOML2* as well
since ${\mathcal{O}}${\rm 7} will let Eq.~(\ref{eq:woml1}) through but 
will let through neither the orthomodularity, nor Eq.~(\ref{eq:woml2}),
and ${\mathcal{O}}${\rm 8} will let through neither the
orthomodularity, nor Eq.~(\ref{eq:woml2}), nor Eq.~(\ref{eq:woml1}).

${\mathcal{O}}${\rm 6} will also let us refine the equivalence class
used for the completeness proof of $\mathcal{CL}$, so that the
Lindenbaum-Tarski algebras be O6 and WDL*.

To obtain OML and DL Lindenbaum algebras we make use of the standard
equivalence classes without the conjoined terms.

All these equivalence classes are relations of congruence. 

\begin{theorem}\label{th:congruence-nonoml}
The relations of {\em equivalence} 
$\approx_{\Gamma,\mathcal{QL},{\rm i}}$, ${\rm i=6,7,8}$, or simply
$\approx_{\rm i}$, ${\rm i=6,7,8}$, defined as
\begin{eqnarray}\label{eq:equiv-oml-re} 
\hskip-5pt A&\approx_{\rm i}&B\hskip8pt{\buildrel\rm
def\over =}\
\Gamma\vdash
A\equiv_q B\ \&\ (\forall o_{\rm i}\in{\mathcal{O}}{\rm i}) \nonumber\\
&&[(\forall X\in\Gamma)(o_{\rm i}(X)=\textstyle{1})
\Rightarrow o_{\rm i}(A)=o_{\rm i}(B)] \nonumber\\
&&{\rm i=6,7,8}.
\end{eqnarray}
are relations of congruence, where\/ 
$\Gamma\subseteq{\mathcal{F}}^\circ$.
\end{theorem}

{\parindent=0pt{\em Proof. }
Let us first prove that $\approx$ is an equivalence
relation. $\>A\approx A\>$ follows from A1
[Eq.~(\ref{eq:kalmb-A1})] of system $\mathcal{QL}$ and the identity
law of equality.
If $\Gamma\vdash A\equiv B$, we can detach the left-hand
side of A12 to conclude $\Gamma\vdash B\equiv A$, through the use of
A13 and repeated uses of A14 and R1.  From this and commutativity
of equality, we conclude $\>A\approx B\>\Rightarrow\>
B\approx A$.  (For brevity we will mostly not mention further uses 
of A12, A13, A14, and R1 in what follows.)
 The proof of transitivity runs as follows (${\rm i}=6,7,8$).
\begin{eqnarray}
A\approx B&&\quad\&\quad B\approx C\label{line1}\\
&&\Rightarrow\ \Gamma\vdash A\equiv   B\quad\&\quad \Gamma\vdash
B\equiv   C\nonumber\\
&&\hskip-20pt\&\ (\forall o\in{\mathcal{O}}{\rm i})
[(\forall X\in\Gamma)(o(X)=1)\ \Rightarrow\
o(A)=o(B)]\nonumber\\
&&\hskip-20pt\&\ (\forall o\in{\mathcal{O}}{\rm i})
[(\forall X\in\Gamma)(o(X)=1)\ \Rightarrow\
o(B)=o(C)]\nonumber\\
&&\Rightarrow\ \Gamma\vdash A\equiv   C \nonumber\\
&&\hskip-20pt\&\ (\forall o\in{\mathcal{O}}{\rm i})[(\forall
X\in\Gamma)(o(X)=1)\nonumber\\
&&\qquad\qquad\qquad\Rightarrow o(A)=o(B)\ \&\ o(B)=o(C)]\nonumber\\
&&\Rightarrow\ \Gamma\vdash A\equiv   C \nonumber\\
&&\hskip-20pt\&\ (\forall o\in{\mathcal{O}}{\rm i})
[(\forall X\in\Gamma)(o(X)=1)\ \Rightarrow\
o(A)=o(C)]\nonumber\\
&&\Rightarrow\ A\approx C\nonumber
\end{eqnarray}
$\Gamma\vdash A\equiv C$ above follows from A2 and the 
metaconjunction in the second but last line reduces to  
$\ o(A)=o(C)\ $ by transitivity of equality.

In order to be a relation of congruence, the relation of
equivalence must be compatible with the operations $\neg$ and
$\vee$. These proofs run as follows (${\rm i}=6,7,8$).
\begin{eqnarray}
A\approx B&&\label{line-1}\\
&&\Rightarrow\Gamma\vdash A\equiv B\nonumber\\
&&\hskip-30pt\&\ \ (\forall o\in{\mathcal{O}}{\rm i})
[(\forall X\in\Gamma)(o(X)=1)\ \Rightarrow\ o(A)=o(B)]\nonumber\\
&&\Rightarrow\Gamma\vdash\neg A\equiv\neg B\nonumber\\
&&\hskip-30pt\&\ \ (\forall o\in{\mathcal{O}}{\rm i})
[(\forall X\in\Gamma)(o(X)=1)\ \Rightarrow\ o(A)'=o(B)']
\nonumber\\
&&\Rightarrow\Gamma\vdash\neg A\equiv\neg B\nonumber\\
&&\hskip-30pt\&\ \ (\forall o\in{\mathcal{O}}{\rm i})
[(\forall X\in\Gamma)(o(X)=1)\ \Rightarrow\ o(\neg A)=o(\neg B)]
\nonumber\\
&&\Rightarrow\neg A\approx\neg B\nonumber
\end{eqnarray}
\begin{eqnarray}
A\approx B&&\label{line-11}\\
&&\Rightarrow \Gamma\vdash A\equiv B\nonumber\\
&&\hskip-20pt\&\ \ (\forall o\in{\mathcal{O}}{\rm i})
[(\forall X\in\Gamma)(o(X)=1)\ \Rightarrow\ o(A)=o(B)]\nonumber\\
&&\Rightarrow\Gamma\vdash(A\vee C)\equiv(B\vee C)\nonumber\\
&&\hskip-30pt\&\ \ (\forall o\in{\mathcal{O}}{\rm i})
[(\forall X\in\Gamma)(o(X)=1)\nonumber\\
&&\Rightarrow\ o(A)\cup o(C)=o(B)\cup o(C)]\nonumber\\
&&\Rightarrow(A\vee C)\approx(B\vee C)\nonumber
\end{eqnarray}
In the second step of Eq.~\ref{line-1}, we used A3.  In the
second step of Eq.~\ref{line-11}, we used A4 and A10.
For the quantified part of these expressions, we applied the definition
of ${\mathcal{O}}{\rm i}$, ${\rm i}=6,7,8$.
$\hfill\blacksquare$}

\begin{theorem}\label{th:congruence-oml}
The relation of {\em equivalence} 
$\approx_{\Gamma,\mathcal{QL},{\rm 1}}$, or simply
$\approx_{\rm 1}$, defined as
\begin{eqnarray}\label{eq:equiv-oml-re-q} 
\hskip-5pt A\approx_{\rm 1} B\hskip8pt{\buildrel\rm
def\over =}\ \Gamma\vdash A\equiv_q B
\end{eqnarray}
is a relation of congruence, where\/ 
$\Gamma\subseteq{\mathcal{F}}^\circ$.
\end{theorem}

{\parindent=0pt{\em Proof. }
The proof for the relation of equivalence given by 
Eq.~(\ref{eq:equiv-oml-re-q}) is the well-known standard one.  
$\hfill\blacksquare$}

\begin{theorem}\label{th:congruence-nondl}
The relation of {\em equivalence} 
$\approx_{\Gamma,\mathcal{CL},{\rm 6}}$, or simply
$\approx_{\rm\bar{6}}$, defined as
\begin{eqnarray}\label{eq:equiv-noml-re-cl6} 
\hskip-5ptA&\approx_{\rm\bar{6}}&B\hskip8pt{\buildrel\rm
def\over =}\
\Gamma\vdash
A\equiv_c B\ \&\ (\forall o_{\rm 6}\in{\mathcal{O}}{\rm i}) \nonumber\\
&&[(\forall X\in\Gamma)(o_{\rm 6}(X)=\textstyle{1})
\Rightarrow o_{\rm 6}(A)=o_{\rm 6}(B)]
\end{eqnarray}
is a relation of congruence, where\/ 
$\Gamma\subseteq{\mathcal{F}}^\circ$.
\end{theorem}

{\parindent=0pt{\em Proof. }
As given in \cite{pm-ql-l-hql1}$\hfill\blacksquare$}

\begin{theorem}\label{th:congruence-dl}
The relation of {\em equivalence}
$\approx_{\Gamma,\mathcal{CL},{\rm 2}}$, or simply 
$\approx_{\rm 2}$,
defined as
\begin{eqnarray}\label{eq:equiv-oml-re-dl} 
\hskip-5ptA\approx_{\rm 2} B\hskip8pt{\buildrel\rm
def\over =}\
\Gamma\vdash
A\equiv_c B
\end{eqnarray}
is a relation of congruence, where\/ 
$\Gamma\subseteq{\mathcal{F}}^\circ$.
\end{theorem}

{\parindent=0pt{\em Proof. }
The proof for the relation of equivalence given by 
Eq.~(\ref{eq:equiv-oml-re-dl}) is the well-known standard one.  
$\hfill\blacksquare$}

\begin{definition}\label{D:equiv-class-sets-woml}
The equivalence classes for a wff $A$ under the relation of 
equivalence $\approx$ given by 
Eqs.~{\em (\ref{eq:equiv-oml-re}), (\ref{eq:equiv-oml-re-q}), 
(\ref{eq:equiv-noml-re-cl6})}, and {\em (\ref{eq:equiv-oml-re-dl})} are
defined as $|A|=\{B\in {\mathcal{F}}^\circ:A\approx B\}$, and we denote
${\mathcal{F}}^\circ/\!\approx\ =\{|A|:A\in {\mathcal{F}}^\circ\}$.
The equivalence classes define the natural morphism
$f:{\mathcal{F}}^\circ\longrightarrow
{\mathcal{F}}^\circ/\!\approx$, which gives
$f(A)\ =^{\rm def}\ |A|$. We write $a=f(A)$, $b=f(B)$, etc.
\end{definition}

\begin{lemma}\label{L:equality-non-q}
The relation $a=b$ on ${\mathcal{F}}^\circ/\!\approx$ is given by:
\begin{eqnarray}
|A|=|B|\qquad&\Leftrightarrow&\qquad A\approx B
\label{eq:equation-non-om-q}
\end{eqnarray}
\end{lemma}

\begin{lemma}\label{L:lind-alg-non-q} The Lindenbaum-Tarski algebras
${\mathcal{A}}_j=\langle {\mathcal{F}}^\circ/\!\approx_j,\neg/
\!\approx_j,\vee/\!\approx_j\rangle$, $j={\rm 6,7,8,1,{\bar{6}},2}$
are {\rm WOML*} (or {\rm O6}), or {\rm WOML1*}, or {\rm WOML2*}, or 
{\rm OML}, or {\rm WDL*} (or {\rm O6}), or {\rm DL}, i.e., 
Eqs.~(\ref{eq:notnot})--(\ref{eq:aAbc}) and (\ref{eq:woml-a}), or
(\ref{eq:woml1}), or (\ref{eq:woml2}), or (\ref{eq:qm-as-id-q}), 
or (\ref{eq:wdl-a3}), or (\ref{eq:qm-as-id-c})
hold for $\neg/\!\approx_j$ and $\vee/\!\approx_j$, 
$j={\rm 6,7,8,1,{\bar{6}},2}$, as  $'$ and $\cup$, respectively. 
\end{lemma}

{\parindent=0pt{\em Proof. }
To prove the $\Gamma\vdash A\equiv B$ part of the $A\approx B$
definition, we prove of the ortholattice conditions,
Eqs.~(\ref{eq:notnot})--(\ref{eq:aAbc}), from A9, the dual of A7, the
dual of A5, etc., analogous to the similar proofs in
\cite{pm-ql-l-hql1} and \cite{pmjlc08}). For Eqs.~(\ref{eq:woml1}) and
(\ref{eq:woml2}) we use Lemma 3.5 from Ref.~\cite{mpcommp99} according
to which any $t=1$ condition that holds in OML also holds in any WOML.
Program {\tt beran} \cite{mpijtp03b} shows that the expressions
$((a\to_1b)\equiv(b\to_1a))\equiv(a\equiv b)$ and 
$((a\equiv b)'\to_1a')\equiv(a\to_1b)$ reduce to 1 in an OML. 
By the aforementioned Lemma 3.5 this means that 
$((a\to_1b)\equiv(b\to_1a))\equiv(a\equiv b)=1$ and 
$((a\equiv b)'\to_1a')\equiv(a\to_1b)=1$ in any WOML. 
Now the $\Gamma\vdash A\equiv B$ part from 
Eq.~(\ref{eq:equiv-oml-re}) forces these WOML conditions 
into Eqs.~(\ref{eq:woml1}) and (\ref{eq:woml2}).
For the quantified part of the $A\approx B$ definition,
lattice O6 is a (proper) WOML. For the OML, we carry out the 
proof with the relation of equivalence without the 
quantified part in Eq.~(\ref{eq:equiv-oml-re}). Then the 
$\Gamma\vdash A\equiv B$ part from Eq.~(\ref{eq:equiv-oml-re})
forces the condition $(a\cup(a'\cap(a\cup b)))\equiv (a\cup b)=1$
which holds in any ortholattice into the OM law given by 
Eq.~(\ref{eq:qm-as-id-q}).
$\phantom .$\hfill$\blacksquare$}

We stress here that the Lindenbaum-Tarski algebras 
${\mathcal{A}}_j$, $j=6,7,8,{\bar{6}}$ from Lemma 
\ref{L:lind-alg-non-q} will be uniquely assigned to 
$\mathcal{QL}$ and $\mathcal{CL}$ via Theorems
\ref{th:completeness-woml} and \ref{th:completeness-wdl} in the sense
that we have to use the relations of congruence given by 
Eqs.~(\ref{eq:equiv-oml-re},\ref{eq:equiv-noml-re-cl6}) and 
that we cannot use those given by 
Eqs.~(\ref{eq:equiv-oml-re-q},\ref{eq:equiv-oml-re-dl}).
For ${\mathcal{A}}_j$, $j=1,2$ we have to use the latter ones and 
we cannot use the former ones. This is in contrast to the 
completeness proof given in \cite{mpcommp99} where we 
did not consider the set-theoretical difference WOML* but 
only WOML. But since WOML contains OML (unlike WOML*), 
in \cite{mpcommp99} (unlike in this paper), in \cite{mpcommp99} we
could have used both relations of congruence (\ref{eq:equiv-oml-re})
and (\ref{eq:equiv-oml-re-q}) to prove the completeness. Here, with
WOML* we can only use (\ref{eq:equiv-oml-re}). We see that the usage of 
set-theoretical differences in this paper establishes a correlation
between lattice models and equivalence relations for a considered logic
as shown in Fig.~\ref{fig:ql-cl}. 

\begin{lemma}\label{L:lind-alg-eq-1-q}
In the Lindenbaum-Tarski algebra $\mathcal{A}$, if
$f(X)=\textstyle{1}$ for all $X$ in\/ $\Gamma$ implies
$f(A)=\textstyle{1}$,
then\/ $\Gamma\vdash A$.
\end{lemma}

{\parindent=0pt{\em Proof.} Let us assume that
$f(X)=\textstyle{1}$ $X\in\Gamma$ implies $f(A)=\textstyle{1}$ i.e.,
$|A|=\textstyle{1}=|A|\cup|A|'=|A\vee\neg A|$, where the 1st equality
follows from Def.~\ref{D:equiv-class-sets-woml}, the 2nd one from
Eq.~(\ref{eq:onezero}) (the definition of $\textstyle{1}$ in OL)
and the 3rd one from the fact that $\approx$ is a congruence.
Hence $A \approx (A\vee\neg A)$, which means $\Gamma\vdash
A\equiv (A\vee\neg A)\ \&\ (\forall o\in{\mathcal{O}}6)[(\forall
X\in\Gamma)(o(X)=\textstyle{1}) \Rightarrow o(A)=o((A\vee\neg A))]$.
The same holds for $\mathcal{O}7$ and $\mathcal{O}8$. When we drop
the second conjunct, this yields $\Gamma\vdash A\equiv (A\vee\neg A)$.
Now, in any OL, we have $a\equiv (a\cup a')=a$.  By mapping the steps
of a proof of this lattice identity to steps of a proof in the logic,
we prove $\vdash (A\equiv (A\vee\neg A))\equiv A$ from $\mathcal{QL}$
axioms A2--A14. By detaching the left-hand side, with the help of A12,
A13, A14, and R1, we arrive at $\Gamma\vdash A$.
$\phantom .$\hfill$\blacksquare$}

\begin{theorem}\label{th:non-orthomodular}The orthomodular law 
does not hold in $\mathcal{A}_j$, $j={\rm 6,7,8,}$ for models 
{\rm WOML*} ({\rm O6}), {\rm WOML1*}, and {\rm WOML2*}.
\end{theorem}

{\parindent=0pt{\em Proof. } 
We assume ${\mathcal F}^\circ$ contains at least 2 primitive
propositions $p_0,p_1,\ldots$. Let us pick up a  valuation $o$ that
maps two of them, $A$ and $B$, to distinct nodes $o(A)$ and $o(B)$ of
O6 which are neither $\textstyle{0}$ nor $\textstyle{1}$ such that
$o(A)\le o(B)$, meaning that $o(A)$ and $o(B)$ are on the same side
of O6 in Fig.\ref{fig:womls}]. In O6, as we can see from
Fig.\ref{fig:womls}], we have  $\>o(A)\cup o(B)=o(B)$ and
$o(A)\cup(o(A)'\cap(o(A)\cup o(B)))
=o(A)\cup(o(A)'\cap o(B))= o(A)\cup \textstyle{0}=o(A)$.
Therefore $o(A)\cup o(B)\ne o(A)\cup (o(A)' \cap (o(A)\cup o(B))$,
i.e., $o(A\vee B)\ne o(A\vee (\neg A\wedge(A\vee B)))$.
This falsifies $(A\vee B)\approx (A\vee(\neg A\wedge (A\vee B))$ 
which is actually very the orthomodularity \cite{pav87,pav89}.
Hence, $a\cup b\ne a\cup(a'\cap(a\cup b))$, what amounts to an
counterexample to the orthomodular law for
${\mathcal F}^\circ/\!\approx$. We can follow the steps given above by
taking $o(A)=x$ and $o(B)=y$ in Fig.~\ref{fig:womls}(a). For O7 and O8
the proofs are analogous. For instance, the orthomodularity is violated
in Fig.~\ref{fig:womls}(b) for $o(A)=x$ and $o(B)=y$ and in
Fig.~\ref{fig:womls}(c) for $o(A)=w$ and 
$o(B)=y$. $\phantom .$\hfill$\blacksquare$}

\begin{theorem}\label{th:orthomodular}The orthomodular law holds
in $\mathcal{A}_1$ for an {\em OML} model.
\end{theorem}

{\parindent=0pt{\em Proof. } Well-known.
$\phantom .$\hfill$\blacksquare$}

\begin{theorem}\label{th:non-distr} The distributive law 
does not hold in $\mathcal{A}_{\bar{6}}$, for {\rm WDL*} ({\rm O6}).
\end{theorem}

{\parindent=0pt{\em Proof. } As given in \cite{pm-ql-l-hql1}.
$\phantom .$\hfill$\blacksquare$}

Eric Schechter \cite[Sec.~9.4]{Schechter} gives a set valued 
interpretation to O6 by assigning \{$-1,0,1,$\} to 1 in our 
Fig.~\ref{fig:womls}(a), \{$-1,0$\} to $y$, \{0,1\} to $x'$, 
\{$-1$\} to $x$, \{1\} to $y'$, and $\varnothing$ to 0, and calls 
it the {\em hexagon interpretation\/}. ``The hexagon interpretation is
not distributive. That fact came as a surprise to some logicians, 
since the two-valued logic itself is
distributive.''~\cite[Sec.~9.5]{Schechter} Schechter also 
gives crystal (6 subsets) and Church's diamond (4 subsets) set 
valued interpretations of $\mathcal{CL}$ in his 
Secs.~9.7.-13.~and 9.14.-17.  

\begin{theorem}\label{th:distributive} The distributive law holds
in $\mathcal{A}_2$ for a {\em DL} model (Boolean algebra).
\end{theorem}

{\parindent=0pt{\em Proof. } Well-known.
$\phantom .$\hfill$\blacksquare$}

\begin{lemma}\label{L:model-wdol}$\mathcal{M}_{\mathcal{A}_j}=\langle
\mathcal{A}_j,f\rangle$, $j={\rm 6,7,8,1,{\bar{6}},2}$, is a proper 
{\em WOML*} ({\em O6}), {\em WOML1*}, {\em WOML2*}, {\em OML}, 
{\em WDL*} ({\em O6}), or {\em DL} model.
\end{lemma}

{\parindent=0pt{\em Proof. } Follows from Lemma 
\ref{L:lind-alg-non-q}. $\phantom{....}$\hfill$\blacksquare$}

Now we are able to prove the completeness of $\mathcal{QL}$ and
$\mathcal{CL}$, i.e., that if a formula {\rm A} is a consequence of a 
set of wffs $\Gamma$ in all O6, WOML*, WOML1*, WOML2*, and OML 
models and in all O6, WDL*, and DL models then 
$\Gamma\vdash_{\mathcal{QL}} A$ and $\Gamma\vdash_{\mathcal{CL}} A$,
respectively. In particular, when $\Gamma=\varnothing$,
all valid formulae are provable in $\mathcal{QL}$.  

\begin{theorem}\label{th:completeness-woml}{\rm[Completeness of
quantum logic]}
$$\ \Gamma\vDash_{\mathcal{M}_{{\mathcal A}_j}}A\ \Rightarrow\ 
\Gamma\vdash_{\mathcal{QL}}A,\quad j={\rm 6,7,8,1}.$$
\end{theorem}

{\parindent=0pt{\em Proof. }
$\Gamma\vDash_\mathcal{M} A$ means that in all WOML* (O6), WOML1*, 
WOML2*, and OML  models $\mathcal{M}$, if $f(X)=\textstyle{1}$ for 
all $X$ in $\Gamma$, then $f(A)=\textstyle{1}$ holds. In particular, 
it holds for $\mathcal{M}_\mathcal{A}=\langle\mathcal{A},f\rangle$, 
which is a WOML* (O6), WOML1*, WOML2*, or OML  model by Lemma 
\ref{L:model-wdol}. Therefore, in the Lindenbaum-Tarski algebra 
$\mathcal{A}$, if $f(X)=\textstyle{1}$ for all $X$ in $\Gamma$, 
then $f(A)=\textstyle{1}$ holds. By Lemma \ref{L:lind-alg-eq-1-q}, 
it follows that $\Gamma\vdash A$. $\phantom .$\hfill$\blacksquare$}

\begin{theorem}\label{th:completeness-wdl}{\rm[Completeness of
classical logic]}
$$\ \Gamma\vDash_{\mathcal{M}_{{\mathcal A}_j}}A\ \Rightarrow\ 
\Gamma\vdash_{\mathcal{CL}}A,\quad j={{\bar{6}},2}.$$
\end{theorem}

{\parindent=0pt{\em Proof.} As given in \cite{pm-ql-l-hql1}.
$\phantom .$\hfill$\blacksquare$}

\section{Discussion}
\label{sec:conclusion}

We have shown that quantum and classical axiomatic logics are 
metastructures for dealing with different algebras, in our case 
lattices, as their models. On the one hand, well formed formulas in 
logic can be mapped to equations in different lattices, and on the 
other, equations from one lattice, we are more familiar with, or 
which is simpler, or easier to handle, can be translated into 
equations of another lattice, through the logic which they are 
both models of. 

In Section \ref{sec:models} we proved that quantum logic can be 
modelled by five different lattice models only one of which is 
orthomodular and that classical logic can be modelled by at least 
three lattice models only one of which is distributive. As we 
conjectured in \cite{pmjlc08} and partly confirmed by means of the two
new models (WOML1* and WOML2*) presented in this paper, there might be
many more, possibly infinitely many, different lattice models quantum
and classical axiomatic logics can be modelled with. (See also the
remarks below Theorem \ref{th:non-distr}.)

\begin{figure}[ht]
\center
\includegraphics[width=0.49\textwidth]{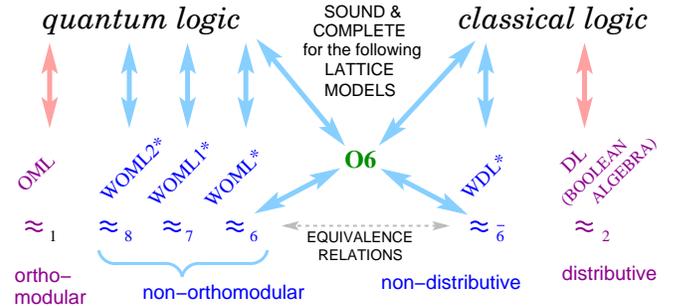}
\caption{Lattice models of quantum and 
classical logics together with the corresponding equivalence 
relations which define their Lindenbaum-Tarski algebras.}
\label{fig:ql-cl}
\end{figure}

The models known to us so far are presented in a chart in
Fig.~\ref{fig:ql-cl}. The key step that allows the multiplicity of
lattice models for both logics is the refinement of the equivalence
relations for the Lindenbaum-Tarski algebras in Theorems 
\ref{th:congruence-nonoml}, \ref{th:congruence-oml}, 
\ref{th:congruence-nondl}, and \ref{th:congruence-dl}. 
They are also given in the chart where we can see that two 
different equivalence relations enable O6 to be a model of both 
quantum and classical logic. This is possible because 
both the weak orthomodularity (\ref{eq:woml-a}) and the 
weak distributivity (\ref{eq:wdl-a4}) pass O6 as pointed out 
below Def.~\ref{def:wdl*}.  

The essence of the equivalence classes of the Lindenbaum-Tarski 
algebras is that they are determined by special simple lattices, 
e.g., those shown in Fig.~\ref{fig:womls}, in which conditions that 
define particular other lattice models, fail. The failure is significant
because it proves that the orthomodularity (\ref{eq:qm-as-id-q}) 
of OML is not needed to prove the completeness of quantum logic for 
WOML2*, that neither the orthomodularity (\ref{eq:qm-as-id-q}), 
nor condition (\ref{eq:woml2}) is needed to prove the completeness 
for WOML1*, and that  neither the orthomodularity 
(\ref{eq:qm-as-id-q}), nor condition (\ref{eq:woml2}), nor condition 
(\ref{eq:woml1}) is needed for  WOML*. 

With today's computational technology we employ only bits and qubits
which correspond to two-valued DL (digital, binary, two-valued Boolean
algebra) and OML, respectively. This means that their possible
valuations are reduced to \{{\tt TRUE,FALSE}\} valuation for classical 
computation, i.e., when we implement classical logic, and to Hasse
diagrams for quantum computation when we implement quantum logic
\cite{bdm-ndm-mp-1}. So, it would be interesting to investigate how
other valuations, i.e., various WOMLs and WDLs, can be implemented
in complex circuits. That would provide us with the possibility of
controlling essentially different algebraic structures (models)
implemented into radically different hardware (logic circuits
consisting of logic gates) by the same logic that we use today with
the standard bit and qubit gate technology.

With these possible applications of quantum and classical 
logics we come back to the question which we started with: 
``Is Logic Empirical?'' We have seen that logic is not 
{\em uniquely\/} empirical since it can simultaneously 
describe distinct realities. However, we have also seen 
(cf.~Fig.~\ref{fig:ql-cl}) that by means of chosen relations 
of equivalence we can link particular kinds of 
``empirical'' models to quantum logic on the one hand and to
classical logic, on the other. Let us therefore briefly review 
the most recent elaborations on the question given by 
Bacciagaluppi \cite{bacc-09} and Baltag and Smets 
\cite{baltag-smets-11}. They state: ``Quantum logic is suitable 
as a logic that locally replaces classical logic when used to 
describe ``a class of propositions in the context of quantum 
mechanical experiments''.''

Our results show that this point can be supported as follows. 
The propositions of quantum logic correspond to elements of a 
Hilbert lattice and are not directly linked to measurement 
values. Such logic employs models which evaluate particular 
combinations of propositions and tell us whether they are true 
or not. Evaluation means mapping from a set of propositions to 
lattice through which a correspondence with measurement values
indirectly emerge. Since the strongest algebra (i.e., not
O6, or WOML*, or WOML1*, or WOML2*, or \dots?) must be an
orthomodular lattice but cannot be a Boolean algebra, we can say
that quantum logic which has an orthomodular lattice as one of its
models is ``empirical'' whenever we theoretically describe quantum
measurements, simply because it {\em can\/} be linked to a model
which serves for such a description: an orthomodular Hilbert lattice,
i.e., the lattice of closed subspaces of a complex Hilbert space. 

\section*{Acknowledgements}
Supports by the Alexander von Humboldt Foundation and the
Croatian Science Foundation through project IP-2014-09-7515 as
well as CEMS funding by the Ministry of Science, Education and
Sports of Croatia are acknowledged. Computational support was
provided by the cluster {\em Isabella} of the {\em University
Computing Centre} of the {\em University of Zagreb} and by the
{\em Croatian National Grid Infrastructure}.

\section*{Conflicts of Interest}
The author declares that there are no conflicts of interest
regarding the publication of this paper.

\end{document}